\documentclass[12pt]{article}
\usepackage{amsmath}
\usepackage{amssymb}
\newcommand{\p}[1]{(\ref{#1})}

\newcounter{rown}
\def\bl{\setcounter{rown}{\value{equation}}
        \stepcounter{rown}\setcounter{equation}0
        \def\theequation{\thesection.\arabic{rown}\alph{equation}}
        }
\def\el{\setcounter{equation}{\value{rown}}
        \def\theequation{\thesection.\arabic{equation}}
        }

\newcommand{\be}{\begin{equation}}\newcommand{\ee}{\end{equation}}
\newcommand{\bea}{\begin{eqnarray}}\newcommand{\eea}{\end{eqnarray}}
\newcommand{\nn}{\nonumber\\[6pt]}
\newcommand{\nns}{\\[10pt]}
\renewcommand{\theequation}{\thesection.\arabic{equation}}

\begin{document}
\thispagestyle{empty}

\title{Higher Spins from Nonlinear Realizations of $OSp(1|8)$}

 \author{Evgeny Ivanov${~}^{1)}$\thanks{eivanov@thsun1.jinr.ru}
 \ and \
 Jerzy Lukierski${~}^{2)\ \dag} $\thanks{lukier@ift.uni.wroc.pl}\thanks{Supported
  by KBN grant 1P03B01828.}
 \\
$^{1)}${\it\small  Bogoliubov Laboratory of Theoretical Physics, JINR,}\\
{\it \small 141980 Dubna, Moscow Region, Russian Federation}\\
$^{2)}${\it\small  Institute for Theoretical Physics, University of
Wroc{\l}aw,} \\
{\it \small Wroc{\l}aw, Poland}\\
}
\date{}
\maketitle
\thispagestyle{empty}

\begin{abstract}

\noindent We exhibit surprising relations between higher spin theory
and nonlinear realizations of the supergroup $OSp(1|8)$, a minimal
superconformal extension of $N=1$, $4D$ supersymmetry with tensorial
charges. We construct a realization of $OSp(1|8)$ on the coset
supermanifold $OSp(1|8)/SL(4,R)$ which involves the tensorial
superspace $R^{(10|4)}$ and Goldstone superfields given on it. The
covariant superfield equation encompassing the component ones for
all integer and half-integer massless higher spins amounts to the
vanishing of covariant spinor derivatives of the suitable Goldstone
superfields, and, via Maurer-Cartan equations, to the vanishing of
$SL(4,R)$ supercurvature in odd directions of $R^{(10|4)}$. Aiming
at higher spin extension of the Ogievetsky-Sokatchev formulation of
$N=1$ supergravity, we generalize the notion of $N=1$ chirality and
construct first examples of invariant superfield actions involving a
non-trivial interaction. Some other potential implications of
$OSp(1|8)$ in the proposed setting are briefly outlined.

\end{abstract}

\newpage

\section{Introduction}
\setcounter{page}{1}

Since the seminal papers by Fradkin and Vasiliev \cite{FV}, the
theory of higher spin fields is under intensive development (see
e.g. \cite{rev,rev2} and refs. therein). Nowadays it attracts vast
attention due to  its profound relations to string theory and
AdS/CFT hypothesis. A concise and suggestive way to deal with higher
spins is to allow for the dependence of fields on additional
coordinates, in particular the tensorial ones, generated by
tensorial charges \cite{ill4}--\cite{ill6}. In $4D$, in order to
justify geometrically the appearance of the tensorial charges, one
should extend the standard supersymmetry algebra to $4D$ counterpart
of $M$-theory algebra \cite{tnm,fpn} and look for the corresponding
dynamical models.

Developing the conjecture of Fronsdal \cite{ill4}, Vasiliev has
shown in \cite{ill5} that the free $4D$ higher spin field theory can
be described by the pair of bosonic and fermionic fields $b(Y),
f_{\hat{\alpha}} (Y)$ ($\hat{\alpha} = 1,2,3,4$) defined on
ten-dimensional real tensorial space
  \begin{equation}\label{ll1}
    Y^{\hat\alpha \hat\beta}=
    Y^{\hat\beta\hat\alpha} = \frac{1}{2}\
    x^{m}(\gamma_m)^{\hat\alpha \hat\beta} +
    \frac{1}{4}\ y^{[m \, n]}
    (\gamma_{[m\, n]})^{\hat\alpha\hat\beta} \,
\end{equation}
where $x^m$ are Minkowski space coordinates. A nice superfield form
of these equations, in the tensorial superspace $R^{(10|4)} =
(Y^{\hat\alpha\hat\beta} , \theta^{\hat\alpha}) \equiv
\tilde{Y}^M\,$, was recently suggested
in \cite{ill6}.\footnote{This superfield equation can be also
recovered as a result of quantization of free twistor superparticle
propagating in $R^{(10|4)}$ \cite{ill3}.}

Keeping in mind the distinguished role of tensorial (super)spaces
in the higher spin theory, it is of urgent importance to get better insights
into their geometry, as well as to work out
the superfield methods of the
appropriate model-building, including construction of the
appropriate off-shell superfield actions for higher spins.

The basic aim of the present paper is to show that an adequate
framework for addressing these and related physically motivated
problems is provided by nonlinear realizations of the supergroup
$OSp(1|8)$ which is a minimal superconformal extension of
$N=1,\, 4D$ supersymmetry with tensorial charges \cite{ill8,ill7}.

We construct a nonlinear realization of $OSp(1|8)$ in the supercoset
\begin{equation}\label{ll6}
\widetilde{{\cal K}} = \
    \frac{OSp(1|8)}{SL(4,R)}
\end{equation}
which is the direct analog of the well known coset of the standard
$4D$, $N=1$ superconformal group \footnote{Nonlinear realizations of
$SU(2,2|1)$ in such a coset were considered in \cite{J} as a natural
extension of standard nonlinear realizations of the conformal group
$SO(2,4)$ \cite{Conf}.}\be {\cal K} = \frac{SU(2,2|1)}{SL(2,C)\times
U(1)}\,. \ee  The supercoset \p{ll6} involves as its
parameters the $R^{(10|4)}$ superspace coordinates and some
Goldstone superfields defined on this superspace. Our main tool is
the formalism of left-covariant Cartan one-forms, supplemented with
covariant constraints on the Goldstone superfields covariant
derivatives. These constraints contain, as an essential part, the
inverse Higgs conditions \cite{ill12} allowing one to algebraically
eliminate all Goldstone superfields in terms of single superfield
associated with the dilatation generator. Simultaneously, they imply
the correct dynamical equations for this basic superfield which
coincide, after some field redefinition, with the equation given in
\cite{ill6}. Thus one of the novel points of our nonlinear
realization approach is that the basic superfield encompassing all
higher spins  appears as a parameter of the supercoset
\p{ll6}. Another point is the new geometric interpretation of this
equation. It proves to be the condition of vanishing of the
covariant spinor derivatives of the Goldstone superfields associated
with the generators of dilatations and conformal supersymmetry. Via
Maurer-Cartan equations, these conditions lead to the vanishing
of the $SL(4,R)$ supercurvature along the pure odd directions
in $R^{(10|4)}\,$.

Besides offering a novel view on the free higher spin dynamics in
the superspace $R^{(10|4)}$, the nonlinear realizations approach
allows one to find out another interesting coset supermanifold of
$OSp(1|8)$, which is a generalization of the chiral $N=1, 4D$
superspace. The latter is known to play the fundamental role in
ordinary $N=1$ supersymmetric theories, so its tensorial counterpart
is expected to have similar implications in higher spin $N=1$ theories.
It is $C^{(11|2)} = (x_L^{\alpha\dot\beta)},
z^{\alpha\beta}_L, f_L^{\alpha\dot\beta}, \theta^\alpha_L)$
involving, besides complex Minkowski coordinate and chiral half of
Grassmann coordinates\footnote{We use here $4D$ Weyl spinor
notation.}, also the holomorphic half of the tensorial coordinates
$z_L^{\alpha\beta}$ and the extra complex coordinates
$f_L^{\alpha\dot\beta}$ which provide a holomorphic parametrization
of the `harmonic' coset $SL(4,R)/GL(2,C)$. We define corresponding
generalized chiral superfields and construct for them two $OSp(1|8)$
invariant off-shell actions which are analogs of the kinetic and
potential terms of the ordinary chiral $N=1$ superfields.

The problem of extension of the nonlinear realizations framework to
the non-flat (in particular, corresponding to AdS structure)
tensorial (super)spaces is now under investigation. We hope that our
nonlinear realization approach will prove useful in constructing
off-shell actions for higher spin fields, as well as for better
understanding of the structure of the higher-spin extensions of
superfield $N=1$ supergravity \cite{ill6}. The generalized chirality
seems to be especially promising in the latter aspect, recalling the
Ogievetsky-Sokatchev formulation of $N=1, 4D$ supergravity \cite{OS}.

\setcounter{equation}{0}

\section{$OSp(1|8)$ as a generalized superconformal group}

The even (bosonic) sector of the superalgebra $osp(1|8)$ is the
generalized $4D$ conformal algebra $sp(8)$ which is a closure of
the standard conformal algebra $so(2,4)$ and the algebra $sl(4,R)$.

The algebra $so(2,4)\simeq su(2,2)$ is spanned by the generators
($L_{\alpha \dot{\beta}}, {\overline{L}}_{\dot{\alpha}\dot{\beta}},
P_{\alpha \dot{\beta}}, K_{\alpha\dot{\beta}}, D$) \bl
\begin{eqnarray}
 \label{ll1.1a}
&&
 \left[  P_{\alpha \dot{\beta}}, P_{\gamma \dot{\delta}}\right]
 =
 \left[  K_{\alpha \dot{\beta}}, K_{\gamma \dot{\delta}}\right]
 = 0
 \\
 \label{ll1.1b}
&& [P_{\alpha\dot\beta}, K_{\rho\dot\lambda}] = \frac 12
\left(\epsilon_{\alpha\rho}\bar L_{\dot\beta\dot\lambda}
- \epsilon_{\dot\beta\dot\lambda} L_{\alpha\rho}\right)
-i \epsilon_{\alpha\rho}\epsilon_{\dot\beta\dot\lambda} D\,, \label{Conf}
\\
\label{ll1.1c}
&& [ L_{\alpha\beta},  L_{\rho\lambda}]
= \epsilon_{\alpha\rho} L_{\beta\lambda} + \epsilon_{\beta\rho} L_{\alpha\lambda}
+ \epsilon_{\alpha\lambda} L_{\beta\rho} + \epsilon_{\beta\lambda} L_{\rho\alpha}\,,
\\
\label{ll1.1d}
&& [ L_{\alpha\beta}, P_{\rho\dot\rho}] = \epsilon_{\alpha\rho}P_{\beta\dot\rho}
+ \epsilon_{\beta\rho}P_{\alpha\dot\rho}\,,
\ [ L_{\alpha\beta}, K_{\rho\dot\rho}] = \epsilon_{\alpha\rho}K_{\beta\dot\rho}
+ \epsilon_{\beta\rho}K_{\alpha\dot\rho}\,,
\\
&&[D, P_{\alpha\dot\alpha}] = i P_{\alpha\dot\alpha}\,,
 \quad [D, K_{\alpha\dot\alpha}] = -i K_{\alpha\dot\alpha}\,.
\label{ll1.1e}
\end{eqnarray}
\el The rest of non-vanishing commutators can be obtained by complex
conjugation.

The algebra $sl(4,R)$ is spanned by the generators $(L_{\alpha
\beta}, \overline{L}_{\dot{\alpha}\dot{\beta}}, A, F_{\alpha
\dot{\beta}}, \overline{F}_{\alpha\dot{\beta}})$. The extra
generators $A$, $F_{\rho \dot{\tau}}$,
$\overline{F}_{\rho\dot{\tau}}\equiv (F_{\tau\dot\rho})^*\, $
satisfy the relations \bl \bea \label{ll2.1a} && [
F_{\alpha\dot\beta}, \overline{F}_{\beta\dot\nu} ]
 = 2 \epsilon_{\alpha\beta}\epsilon_{\dot\beta\dot\nu} A
+ 2 \left(\epsilon_{\alpha\beta}\bar L_{\dot\beta\dot\nu}
- \epsilon_{\dot\beta\dot\nu} L_{\alpha\beta}\right)\,,
\\
\label{ll2.1b}
&& [ F_{\alpha\dot\alpha}, F_{\beta\dot\nu} ]
= [ \overline{F}_{\alpha\dot\alpha}, \overline{F}_{\beta\dot\nu} ] = 0\,,
\\
\label{ll2.1c} && [ A, F_{\alpha\dot\beta}] = 2
F_{\alpha\dot\beta}\,, \quad [ A,  \overline{F}_{\alpha\dot\beta}] =
- 2  \overline{F}_{\alpha\dot\beta}\,.\label{sl4R} \eea
\el

The generalized $4D$ conformal algebra $sp(8)$ is a closure of the
algebras $so(2,4)$ and $sl(4,R)\,$. It is obtained by adding to the
generators of $sl(4,R)$ and the vectorial Abelian translation
generators ($P_{\alpha \dot{\beta}}, K_{\alpha \dot{\beta}}$) the
following additional 12 Abelian generators

- ($Z_{\alpha {\beta}}, \overline{Z}_{\dot{\alpha}\dot{\beta}}$)
describing six standard tensorial translations

- (${\widetilde{Z}}_{\alpha \beta},
\overline{{\widetilde{Z}}}_{\dot{\alpha}\dot{\beta}}$) describing six
 conformal tensorial translations.

They satisfy the following commutation relations: \bl
 \bea
 \label{ll2.3a}
   &&
[Z_{\alpha\beta}, {\widetilde{Z}}_{\rho\lambda}]
= \frac 12 \left(\epsilon_{\alpha\rho}
 L_{\beta\lambda} + \epsilon_{\beta\rho} L_{\alpha\lambda} +
\epsilon_{\alpha\lambda} L_{\beta\rho} +
 \epsilon_{\beta\lambda} L_{\alpha\rho} \right) \nonumber \\
&& + \,
\left(\epsilon_{\alpha\rho}\epsilon_{\beta\lambda}
+ \epsilon_{\beta\rho}\epsilon_{\alpha\lambda}\right)
\left(iD - \frac 12 A\right),
\label{MT} \\
 \label{ll2.3b}
&& [P_{\alpha\dot\beta}, {\widetilde{Z}}_{\rho\beta}]
 = \frac 12 \left(\epsilon_{\alpha\rho} \overline{F}_{\beta\dot\beta} +
\epsilon_{\alpha\beta} \overline{F}_{\rho\dot\beta}\right),
[K_{\alpha\dot\beta}, Z_{\rho\beta}]
= \frac 12 \left(\epsilon_{\alpha\rho}F_{\beta\dot\beta} +
\epsilon_{\alpha\beta}F_{\rho\dot\beta}\right),
\\
 \label{ll2.3c}
&&
\nonumber\\
&& [P_{\alpha\dot\alpha}, F_{\beta\dot\beta}]
 = -2 \epsilon_{\dot\alpha\dot\beta} Z_{\alpha\beta}\,,
[Z_{\alpha\beta}, F_{\rho\dot\nu}] = 0\,,   [Z_{\alpha\rho},
\bar F_{\gamma\dot\nu}] =
2\left(\epsilon_{\alpha\gamma} P_{\rho\dot\nu}
 + \epsilon_{\rho\gamma} P_{\alpha\dot\nu} \right),
 \label{ll2.3d} \\
&&
[K_{\alpha\dot\alpha}, F_{\beta\dot\beta}]
 = 2 \epsilon_{\alpha\beta} \bar {\widetilde{Z}}_{\dot\alpha\dot\beta}\,,
[{\widetilde{Z}}_{\alpha\beta},  \overline{F}_{\rho\dot\nu}] = 0\,,
  [{\widetilde{Z}}_{\alpha\rho}, F_{\gamma\dot\nu}] =
2\left(\epsilon_{\alpha\gamma} K_{\rho\dot\nu} +
 \epsilon_{\rho\gamma} K_{\alpha\dot\nu} \right),
 \label{ll2.3e} \\
&& [ A, Z_{\alpha\rho}] = 2 Z_{\alpha\rho}\,, \quad
[ D, Z_{\alpha\rho}] = i Z_{\alpha\rho}\,,
 \label{PM}
 \label{ll2.3f} \\
&& [ A, {\widetilde{Z}}_{\alpha\rho}] = -2
{\widetilde{Z}}_{\alpha\rho}\,, \quad [ D,
{\widetilde{Z}}_{\alpha\rho}] = -i {\widetilde{Z}}_{\alpha\rho}\,.
\label{KT} \eea \el The remaining commutators are either vanishing,
or can be obtained by complex conjugation from the above ones,
taking into account the rules $\overline{A} = A, \overline{D} = D$.

The odd (fermionic) sector of $osp(1|8)$ involves $N=1$ super
 Poincar\'e generators $Q_\alpha, \bar Q_{\dot\alpha}$
and the generators $S_\alpha, \bar S_{\dot\alpha}$ of
 conformal supersymmetry.\footnote{Standard $N=1$, $D = 4$
 superconformal symmetry
$su(2,2|1)$ is not a subalgebra of $osp(1|8)$.
These two superalgebras describe two
different superextensions of $4D$ conformal symmetry
(see discussion in \cite{bils}).}
The basic algebraic relations look as follows

i) basic superalgebra relations \bl \bea \label{llnew2.13a}
&&\{Q_\alpha, \bar Q_{\dot\alpha}\} = 2 P_{\alpha\dot\alpha}\,,
 \quad
\{Q_\alpha, Q_{\beta}\} = 2 Z_{\alpha\beta}\,, \quad
\{\bar Q_{\dot\alpha}, \bar Q_{\dot\beta}\}
= 2 \overline{Z}_{\dot\alpha\dot\beta}\,,\nn
&&
\{S_\alpha, \bar S_{\dot\alpha}\} = 2 K_{\alpha\dot\alpha}\,,
\quad
\{S_\alpha, S_{\beta}\} = 2 {\widetilde{Z}}_{\alpha\beta}\,, \quad
\{\bar S_{\dot\alpha}, \bar S_{\dot\beta}\} =
2 {\overline{\widetilde{Z}}}_{\dot\alpha\dot\beta}\,,\nn
&&
\{Q_\alpha, \bar S_{\dot\beta} \} = F_{\alpha\dot\beta}\,, \quad
\{S_\alpha, \bar Q_{\dot\beta} \} = \bar F_{\alpha\dot\beta}\,, \nn
&&
\{Q_\alpha, S_\beta \} = \epsilon_{\alpha\beta}\left(iD -\frac 12 A \right)
+ L_{\alpha\beta}\,, \nn
&& \{\bar Q_{\dot\alpha}, \bar S_{\dot\beta} \}
 = -\epsilon_{\dot\alpha\dot\beta}\left(iD +\frac 12 A \right) +
\bar L_{\dot\alpha\dot\beta}\,,
\eea

ii) covariance relations for supercharges
\bea \label{llnew2.13v}
&&[A, Q_\alpha ] = Q_\alpha\,, \;\;[A, \bar Q_{\dot\alpha}]
 = - \bar Q_{\dot\alpha}\,, \;\;
[D, Q_\alpha] = \frac i2 Q_\alpha\,, \;\; [D, \bar Q_{\dot\alpha}]
 = \frac i2 \bar Q_{\dot\alpha}\,,
\nn
&& [A, S_\alpha ] = -S_\alpha\,, \;\;[A, \bar S_{\dot\alpha}]
 = \bar S_{\dot\alpha}\,, \;\;
[D, S_\alpha] = -\frac i2 S_\alpha\,, \;\; [D, \bar S_{\dot\alpha}]
 = -\frac i2 \bar S_{\dot\alpha}\,,
\nn
&&
[Q_\alpha, F_{\rho\dot\beta}] = 0\,, \;\;
[Q_\alpha, \overline{F}_{\rho\dot\beta}]
 = 2\epsilon_{\alpha\rho}\bar Q_{\dot\beta}\,,\;\;
[S_\alpha, \overline{F}_{\rho\dot\beta}] = 0\,, \;\;
[S_\alpha, F_{\rho\dot\beta}] = 2\epsilon_{\alpha\rho}\bar S_{\dot\beta}\,,
\nn
&&
[{\widetilde{Z}}_{\alpha\beta}, Q_\rho] = \epsilon_{\alpha\rho} S_\beta
 + \epsilon_{\beta\rho} S_\alpha\,, \quad
[{\overline{\widetilde{Z}}}_{\dot\alpha\dot\beta}, Q_\rho] = 0\,,
\nn
&&
[Z_{\alpha\beta}, S_\rho] = \epsilon_{\alpha\rho} Q_\beta
 + \epsilon_{\beta\rho} Q_\alpha\,, \quad
[{\overline{Z}}_{\dot\alpha\dot\beta}, S_\rho] = 0\,, \nn &&
[P_{\alpha\dot\alpha}, S_\rho] = \epsilon_{\alpha\rho} \bar
Q_{\dot\alpha}\,, \;\; [K_{\alpha\dot\alpha}, Q_\rho] =
\epsilon_{\alpha\rho} \bar S_{\dot\alpha}\,. \label{Odd} \eea \el
All other (anti)commutators vanish except the complex conjugates of
(\ref{Odd}).

\setcounter{equation}{0}
\section{Nonlinear realizations of $OSp(1|8)$}
Before constructing nonlinear realization of $OSp(1|8)$ in the
supercoset \p{ll6}, we consider the bosonic limit of this realization.
Namely, we consider $Sp(8) \subset OSp(1|8)$ and
construct an $Sp(8)$ analog of the nonlinear realization of ordinary
conformal group $SO(2,4)$ in the coset $SO(2,4)/SO(1,3)$ \cite{Conf}.
It corresponds to the choice of the coset $K= Sp(8)/SL(4,R)$ spanned by the
following generators
\begin{equation}\label{llnew3.1}
    K: \ (P_{\alpha \dot{\beta}},  Z_{\alpha\beta },
    {\overline{Z}}_{\dot{\alpha}\dot{\beta}},
K_{ {\alpha} \dot{\beta} },
{\widetilde{Z}}_{ {\alpha}{\beta} },
{\overline{{\widetilde{Z}}}}_{\dot{\alpha}\dot{\beta}}, D
    )\,.
\end{equation}
We represent the coset $K$ by the following element of $Sp(8)$
\be g
= e^{i(x\cdot P + z\cdot {Z})}\,e^{i\phi D} \,e^{i(k\cdot K + t\cdot
{\widetilde{Z}})}
 \label{param}
\ee
where
\be
\label{llnew3.3}
 x\cdot P =
x^{\alpha\dot\alpha}P_{\alpha\dot\alpha}\,, \;k\cdot K =
k^{\alpha\dot\alpha}K_{\alpha\dot\alpha}\,, \; z\cdot {Z} =
z^{\alpha\beta}{Z}_{\alpha\beta} + \bar z^{\dot\alpha\dot\beta}
{\overline{Z}}_{\dot\alpha\dot\beta}\,, \;\; \mbox{etc}\,.
\ee
The group $Sp(8)$ acts on this element from the left, producing the corresponding
transformation of the coset parameters.

According to the general rules of nonlinear realizations,
we are led to consider
$(x^{\alpha\dot\alpha}, z^{\alpha\beta},
\bar z^{\dot\alpha\dot\beta})
 \equiv
  Y^{\hat{\alpha}\hat{\beta}}$
   as coordinates and the rest of
the coset parameters as Goldstone fields living on this extended
ten-dimensional space.
The basic objects of the considered nonlinear realizations
framework are the left-covariant Cartan one-forms
\footnote{Cartan forms
  for the supergroups $OSp(1|n)$
and $OSp(N|n)$ treated as curved versions of tensorial superspaces,
 with AdS subspaces instead of the
Minkowski ones, were constructed in
  \cite{BLPS}- \cite{DV}; for $n=8$ see \cite{J}.
 The new input of our construction is that we treat $OSp(1|8)$ as
  a spontaneously broken symmetry realized in the
 supercoset \p{ll6} in which the ten-dimensional
 $4D$ tensorial superspace
forms a coordinate subspace, while other coset parameters
 are Goldstone superfields with suitable constraints.}
\bea
g^{-1}d g = i(\,\omega_P\cdot P + \omega_Z \cdot
 {{Z}} + \omega_D D
+ \omega_K\cdot K
+ \omega_{\widetilde{Z}}\cdot
 {\widetilde{Z}}
+ \omega_F\cdot F +
\omega_L\cdot L + \omega_A A\,). \label{formsBos} \eea
Explicitly, the forms necessary for our consideration are
\bea && \omega_P^{\alpha\dot\alpha} =
e^\phi\,dx^{\alpha\dot\alpha}\,, \;
\omega_Z^{\alpha\beta} =
e^\phi\,dz^{\alpha\beta}\,, \;\omega_D = d\phi + e^\phi\,(dx\cdot
k) -2e^\phi\,(dz\cdot t)\,,
\nn
&&
\omega_L^{\alpha\beta} = i e^\phi \left[ 2
dz^{\gamma(\alpha}t^{\beta)}_\gamma -\frac{1}{2}
dx^{(\alpha\dot\rho}k^{\beta)}_{\dot\rho} \right], \; \omega_A = i
e^\phi\left[d\bar z^{\dot\alpha\dot\beta}\bar
t_{\dot\alpha\dot\beta} - dz^{\alpha\beta}t_{\alpha\beta}\right],
\nn && \omega_F^{\alpha\dot\beta} =
ie^\phi\left[dz^{\alpha\gamma}k^{\dot\beta}_\gamma -
dx^{\alpha\dot\rho}\bar t^{\dot\beta}_{\dot\rho}\right], \quad
\omega_{\bar{L}}^{\dot\alpha\dot\beta} = \overline{\omega_L^{\alpha\beta}}\,, \quad
\omega_{\bar{F}}^{\dot\alpha\beta}=\overline{\omega_F^{\alpha\dot\beta}}\,.
\label{CartanB}
\eea

Now we are prepared to consider a nonlinear realization of the
supergroup $OSp(1|8)$ in the coset   $\widetilde{\cal K}$ (\ref{ll6}).
For this purpose we  add to the previous coset
generators the  spinor generators, $Q_\alpha, \bar
Q_{\dot\alpha}$ and $S_\alpha, \bar S_{\dot\alpha}$.
Correspondingly, we introduce new coset coordinates, the spinor
coordinates $\theta^\alpha, \bar\theta^{\dot{\alpha}}$ extending the previous
bosonic space $Y^{\hat{\alpha}\hat{\beta}} \equiv (x^{\alpha\dot\beta}, z^{\alpha\beta},
\bar{z}^{\dot\alpha\dot\beta})$ to the superspace
 ${\widetilde{Y}}^M \equiv (Y^{\hat{\alpha}\hat{\beta}}, \theta^\alpha,
\bar\theta^{\dot\alpha})$ and the spinor Goldstone superfields $\psi^\alpha(\widetilde{Y}),
\bar\psi^{\dot\alpha}(\widetilde{Y})$.
  We parametrize the  supercoset elements as follows
  \footnote{We define the contraction of two Weyl
spinors in the standard way, $\psi\cdot\xi =
\psi^\alpha\xi_\alpha\,,\; \bar\psi\bar\xi =
\bar\psi_{\dot\alpha}\cdot\bar\xi^{\dot\alpha}\,,\; (\psi)^2 =
\psi^\alpha\psi_\alpha\,,$ also $ x^2 =
x^{\alpha\dot\alpha}x_{\alpha\dot\alpha}$, etc.} \be
 G = e^{i(\theta Q
+ \bar\theta \bar Q)}\,g\, e^{i(\psi S + \bar\psi \bar S)}\,,\label{suparam}
\ee
where $g$ is the same bosonic coset element as defined in \p{param},
with
all parameter-fields now being superfields on the superspace $\widetilde{Y}^M$. The Cartan
forms are defined by:
\bea G^{-1}dG &=& i (\,
\Omega_Q\cdot Q + \Omega_S\cdot S + \Omega_P\cdot P + \Omega_Z\cdot
Z + \Omega_D D + \Omega_K\cdot
K
\nn &&  + \,
\Omega_{\widetilde{Z}}{}\cdot {\widetilde{Z}} + \Omega_L\cdot L +
\Omega_A A + \Omega_F\cdot F \,) \equiv i \Omega \label{Supforms}
 \eea
 where the notation basically follows the bosonic case and
 $\Omega_Q\cdot Q = \Omega_Q^\alpha Q_\alpha +
\bar\Omega_{Q\dot\alpha}\bar Q^{\dot\alpha}\,$, etc.
Once again, we explicitly present only few forms needed for our purpose
\begin{equation}
\begin{array}{rll}
 &&
  \Omega^\alpha_Q = e^{\frac
12 \phi}\,d\theta^\alpha + i
\hat\omega{}^{\alpha\dot\alpha}_P\bar\psi_{\dot\alpha} + 2i
\hat\omega{}^{\alpha\beta}_{Z} \psi_\beta\,,
\nns
 &&
 \Omega^\alpha_S
= d\psi^\alpha + \frac 12 e^{\frac 12 \phi} (\psi)^2
d\theta^\alpha - e^{\frac 12 \phi} \psi^\alpha (\bar\psi
d\bar\theta) - i e^{\frac 12 \phi}
d\bar\theta_{\dot\alpha}k^{\dot\alpha\alpha}
\nns
 && \qquad +
 2i e^{\frac
12 \phi} d\theta^\beta t^\alpha_\beta + \frac 12 \psi^\alpha
\hat\omega_D + 2i \hat\omega^{\beta\alpha}_L \psi_\beta - i
\psi^\alpha \hat\omega_A + 2i
\hat{\bar\omega}{}^{\alpha\dot\alpha}_F\bar\psi_{\dot\alpha}\,,
\nns
&&
 \Omega^{\alpha\dot\alpha}_P =
\hat\omega{}^{\alpha\dot\alpha}_P\,, \;\;\Omega^{\alpha\beta}_Z =
\hat\omega{}^{\alpha\beta}_Z\,, \;\; \Omega_D= \hat\omega_D +
e^{\frac 12 \phi} d\theta^\alpha\psi_\alpha + e^{\frac 12 \phi}
d\bar\theta_{\dot\alpha}\bar\psi^{\dot\alpha}\,.
\end{array}
\end{equation}
The one-forms with `hat' are obtained from the forms \p{CartanB}
via the replacements
\bea
 &&
dx^{\alpha\dot\alpha} \;\Rightarrow \; \Delta x^{\alpha\dot\alpha}
=dx^{\alpha\dot\alpha} - i\left(\theta^\alpha
d\bar\theta^{\dot\alpha} +
\bar\theta^{\dot\alpha}d\theta^\alpha\right),
\nn &&
dz^{\alpha\beta} \;\Rightarrow \; \Delta z^{\alpha\beta}
=dz^{\alpha\beta} + i \theta^{(\alpha} d\theta^{\beta)}\,, \quad
d\bar z^{\dot\alpha\dot\beta} \;\Rightarrow \; \Delta \bar
z^{\dot\alpha\dot\beta} =d\bar z^{\dot\alpha\dot\beta} +i
\bar\theta^{(\dot\alpha} d\bar\theta^{\dot\beta)}\,. \label{Repl}
\eea

Now let us show that the  fermionic Goldstone superfields
$\psi^\alpha({\widetilde{Y}}),
\bar\psi^{\dot\alpha}({\widetilde{Y}})$, as well as the bosonic ones
$k^{\alpha\dot\alpha}({\widetilde{Y}})$,
$t^{\alpha\beta}({\widetilde{Y}})$ and $\bar
t^{\dot\alpha\dot\beta}({\widetilde{Y}})$, can be covariantly
eliminated by imposing one basic inverse super-Higgs \cite{ill12}
constraint
 \be
\Omega_D = 0\,. \label{IHconstr} \ee
For bosonic Goldstone superfields this
constraint yields \be k_{\alpha\dot\alpha} = -
e^{-\phi}\,\partial_{\alpha\dot\alpha}\phi\,, \;\;t_{\alpha\beta} =
\frac 12 e^{-\phi}\,\partial_{\alpha\beta}\phi\,, \;\; \bar
t_{\dot\alpha\dot\beta} = \frac 12
e^{-\phi}\,\partial_{\dot\alpha\dot\beta}\phi \,, \label{IH} \ee
while for the fermionic ones we obtain \be \psi_\alpha = -e^{-\frac
12 \phi}\,D_\alpha\phi\,, \quad \bar\psi_{\dot\alpha} = -e^{-\frac
12 \phi}\,D_{\dot\alpha}\phi\,, \label{IH2} \ee where \bea &&
D_\alpha = \frac{\partial}{\partial \theta^\alpha} -
i\bar\theta^{\dot\beta}\partial_{\alpha\dot\beta} +i
\theta^\beta\partial_{\alpha\beta}\,, \quad \bar D_{\dot\alpha} =
-\frac{\partial}{\partial \bar\theta^{\dot\alpha}} +
i\theta^{\beta}\partial_{\beta\dot\alpha} - i
\bar\theta^{\dot\beta}\partial_{\dot\alpha\dot\beta}\,,
\label{Ddefin} \\
&& \{D_\alpha, \bar D_{\dot\alpha} \} = 2i
\partial_{\alpha\dot\alpha}\,, \quad \{D_\alpha, D_{\beta} \} = 2i
\partial_{\alpha\beta}\,, \quad \{\bar D_{\dot\alpha}, \bar D_{\dot\beta} \} = 2i
\partial_{\dot\alpha\dot\beta}\,. \label{Dcomm} \eea
Thus all Goldstone superfields have been expressed through the single
basic scalar Goldstone superfield $\phi(\tilde{Y})$ associated with
the dilatonic generator $D$. This superfield is the basic object of the nonlinear
realization considered.

As the last topic of this section we present the transformation
rules of the basic coset parameters under the Poincar\'e and
conformal supersymmetries.

In our case all bosonic transformations are generated in the closure
of the Poincar\'e supersymmetry transformations (left shifts with
$Q_\alpha, \bar Q_{\dot\alpha}$) and transformations
of the `conformal' supersymmetry (left shifts with
$S_\alpha, \bar S_{\dot\alpha}$). So it is enough to know those
transformations which are induced by the left multiplication of the
supercoset element \p{suparam} by an element $e^{i (a\cdot X)}$ with
\be (a\cdot X) = \epsilon^\alpha Q_\alpha +
\bar\epsilon_{\dot\alpha}\bar Q^{\dot\alpha} +\eta^\alpha S_\alpha +
\bar\eta_{\dot\alpha}\bar S^{\dot\alpha}\,.\ee After straightforward
computations using the $osp(1|8)$ structure relations we find \bea
&& \delta x^{\alpha\dot\alpha} =
i(\epsilon^\alpha\bar\theta^{\dot\alpha} +
\bar\epsilon^{\dot\alpha}\theta^\alpha) + 2\eta_\beta
z^{\beta\alpha} \bar\theta^{\dot\alpha} - 2\bar\eta_{\dot\beta} \bar
z^{\dot\beta\dot\alpha} \theta^\alpha - \eta_\beta
x^{\beta\dot\alpha}\theta^\alpha + \bar\eta_{\dot\beta}
x^{\alpha\dot\beta} \bar\theta^{\dot\alpha}\,, \nn && \delta
z^{\alpha\beta} = i\theta^{(\alpha}\epsilon^{\beta)} - 2\eta_\gamma
z^{\gamma(\alpha}\theta^{\beta)} -
\bar\eta_{\dot\alpha}\theta^{(\alpha} x^{\beta)\dot\alpha}\,, \;\;
\delta\bar z^{\dot\alpha\dot\beta} = i
\bar\theta^{(\dot\alpha}\bar\epsilon^{\dot\beta)} +
2\bar\eta_{\dot\gamma} \bar
z^{\dot\gamma(\dot\alpha}\bar\theta^{\dot\beta)} + \eta_\alpha
x^{\alpha(\dot\alpha}\bar\theta^{\dot\beta)}\,, \nn &&
\delta\theta^\alpha = \epsilon^\alpha - 2i \eta_\beta
z^{\beta\alpha} +\frac 12 \,(\theta)^2 \eta^\alpha +
\bar\eta_{\dot\alpha}(\theta^\alpha \bar\theta^{\dot\alpha} -i
x^{\alpha\dot\alpha})\,, \nn && \delta \bar\theta^{\dot\alpha} =
\bar\epsilon^{\dot\alpha} + 2i \bar\eta_{\dot\beta} \bar
z^{\dot\beta\dot\alpha} + \frac 12 (\bar\theta)^2
\bar\eta^{\dot\alpha} +\eta_\alpha
(\theta^\alpha\bar\theta^{\dot\alpha} + i x^{\alpha\dot\alpha})\,,
\label{TranCoord} \\
&& \delta \phi = \phi{}'(\tilde{Y}') - \phi(\tilde{Y}) = -(\eta \theta +
\bar\eta\bar\theta)\,. \label{Tranphi} \eea The transformation
properties of the remaining Goldstone fields can be easily found
using the inverse Higgs expressions \p{IH}, \p{IH2}, the transformation
law \p{Tranphi} and the
transformation rules of the derivatives
$\partial_{\alpha\dot\alpha},
\partial_{\alpha\beta}, \partial_{\dot\alpha\dot\beta}, D_\alpha, \bar D_{\dot\alpha}\,$,
e.g., \bea && \delta
D_\alpha = -(\theta_\alpha\eta^\beta - \eta_\alpha \theta^\beta)
D_\beta - (\eta\theta)D_\alpha
-2(\eta_\alpha\bar\theta^{\dot\beta})\bar D_{\dot\beta}\,, \nn &&
\delta \bar D_{\dot\alpha} =
(\bar\theta_{\dot\alpha}\bar\eta^{\dot\beta} -
\bar\eta_{\dot\alpha}\bar\theta^{\dot\beta})\bar D_{\dot\beta} -
(\bar\eta\bar\theta)\bar D_{\dot\alpha} +
2(\bar\eta_{\dot\alpha}\theta^\beta)D_\beta\,. \label{TransD} \eea

\setcounter{equation}{0}

\section{Higher spin dynamics from Cartan forms}

To see how the higher spin dynamics arises within the nonlinear realizations
approach let us substitute the inverse Higgs expression \p{IH}, \p{IH2} for the
Goldstone superfields into the covariant differentials of the superspace
coordinates, i.e.  $\Omega_Q^\alpha\,, \bar\omega_Q^{\dot\alpha}\,,
\Omega_P^{\alpha\dot\alpha}, \Omega_Z^{\alpha\beta},
\bar\Omega_Z^{\dot\alpha\dot\beta}$, and the covariant
differentials of the fermionic Goldstone superfields
$\Omega_S^{\alpha}, \bar\Omega_S^{\dot\alpha}$.
 As a consequence the fermionic part of
  $\Omega^{\alpha}_{S}$ takes the simple form \be \Omega_S^\alpha | =
\Omega^\beta_Q\left\{e^{-2\phi} D^{[\alpha} D_{\beta]} e^{\phi} \right\}
+\bar\Omega_{\dot\beta Q}\left\{e^{-2\phi}D^{[\alpha} \bar D^{\dot\beta]}
e^{\phi} \right\}. \label{S}
\ee
Then the desired equations for higher
spins follow  from the requirement that  these projections (and their
conjugates) vanish:
 \be (D)^2e^{\phi} = (\bar D)^2 e^{\phi}=0\,, \;\;
\left[D^\alpha, \bar D^{\dot\alpha}\right] e^{\phi} = 0\,.
\label{EQ} \ee
Eqs. \p{EQ} are recognized as the two-component spinor form of the equation suggested in
\cite{ill6} (for $\Phi = e^\phi$). The $OSp(1|8)$ covariance of \p{EQ} can be
directly checked using \p{Tranphi}, \p{TransD}.

We observe that the superfield system \p{EQ}
 amounts to vanishing of
{\it full} covariant {\it spinor} derivatives of the {\it spinor}
Goldstone superfields $\psi^\alpha, \bar\psi_{\dot\alpha}$.
Indeed, by definition \be
 \Omega_S^\alpha | =\Omega^\beta_Q\nabla_\beta \psi^\alpha +
\bar\Omega_{Q \dot\beta}\bar\nabla^{\dot\beta}\psi^\alpha\,, \ee
where, with taking account of \p{IH}, \p{IH2},
\be \nabla_\beta \psi^\alpha =
\delta^\alpha_\beta \,e^{-2\phi}\,(D)^2\,e^{\phi}\,, \quad
\bar\nabla^{\dot\beta}\psi^\alpha = \frac
12\,e^{-2\phi}\left[D^\alpha,  \bar D^{\dot\beta}\right] e^{\phi}\,.
\label{VanCov}
 \ee

Let us present another form of eqs. \p{EQ} which is more
suggestive. Prior to imposing the inverse Higgs constraints, the
covariant derivatives of $\psi^\alpha, \bar\psi_{\dot\alpha}$ are
as follows \bea && \nabla_\beta\psi^\alpha = e^{-\frac 12 \phi}
D_\beta \psi^\alpha -\frac 12 e^{-\frac 12 \phi} \psi^\alpha
D_\beta \phi + 2i t^\alpha_\beta +\frac 12 \delta^\alpha_\beta
(\psi)^2\,, \nn  && \bar\nabla_{\dot\beta}\psi^\alpha = e^{-\frac
12 \phi} \bar D_{\dot\beta} \psi^\alpha -\frac 12 e^{-\frac 12
\phi} \psi^\alpha \bar D_{\dot\beta} \phi  - i
k^\alpha_{\dot\beta} + \psi^\alpha \bar\psi_{\dot\beta}\,.
\label{beforeIH} \eea Then it is easy to show that the dynamical
Eqs. \p{EQ}, as well as the inverse Higgs expressions for the
bosonic Goldstone superfields, can be derived from the following
minimal set of equations \bea && \nabla_\beta\psi^\alpha = 0\,,
\;\; \bar\nabla_{\dot\beta}\psi^\alpha = 0 \;\;\;\mbox{and
c.c.}\,,
\label{Eqcova} \\
&& \nabla_\beta\phi = 0\,, \;\; \bar\nabla_{\dot\beta}\phi = 0\,,
\label{Eqcovb} \eea where $\nabla_\beta\phi\,,
\bar\nabla_{\dot\beta}\phi$ are covariant spinor projections of
the Cartan form $\Omega_D$: \be \nabla_\beta\phi = \psi_\beta +
e^{-\frac 12 \phi} D_\beta \phi\,, \;\;\bar\nabla_{\dot\beta}\phi
= \bar\psi_{\dot\beta} + e^{-\frac 12 \phi} \bar D_{\dot\beta}
\phi\,. \ee Eqs. \p{Eqcovb} express spinor Goldstone superfields
through the superdilaton $\phi$, then Eqs. \p{Eqcova} imply the
expressions \p{IH} for the bosonic Goldstone superfields and
simultaneously yield the dynamical Eqs. \p{EQ}. Actually, it is
the traceless part of the first equation in \p{Eqcova} and the
imaginary part $\sim (\bar\nabla_{\dot\alpha}\psi_\alpha +
\nabla_{\alpha}\bar\psi_{\dot\alpha})$ of the second one which,
together with \p{Eqcovb}, form a kinematical subset in the set
\p{Eqcova}, \p{Eqcovb}. The vanishing of the remaining covariant
projections of the Cartan form $\Omega_D$ (associated with the
forms $\Omega_P^{\alpha\dot\alpha}, \Omega_Z^{\alpha\beta}$ and
$\bar\Omega^{\dot\alpha\dot\beta}_Z$) and, hence, of the whole
$\Omega_D$ (Eq. \p{IHconstr}), is just a consequence of the
Maurer-Cartan equations and the kinematical part of Eqs.
\p{Eqcova}, \p{Eqcovb}.

The formulation based on Eqs. \p{Eqcova}, \p{Eqcovb} is
advantageous also because of its manifest $OSp(1|8)$ covariance
which does not require any explicit checks. Indeed, all $OSp(1|8)$
transformations of the full covariant derivatives have the form of
induced transformations of the stability subgroup $SL(4,R)$ acting
on the spinor indices. The covariance under the $GL(2,C)$
transformations is evident and one should only be convinced of the
covariance under the transformations generated by
$F_{\alpha\dot\alpha}, \bar F_{\alpha\dot\alpha}$. From the
general transformation of the Cartan form \p{Supforms} it is easy
to deduce the transformations of the spinor covariant derivatives
of the involved Goldstone superfields \bea && \delta
(\nabla_\alpha \phi) = -2i
\bar\lambda_{\alpha\dot\alpha}\bar\nabla^{\dot\alpha}\phi\,,
\;\;\delta(\bar\nabla_{\dot\alpha}\phi) = -2i
\lambda^{\alpha}_{\dot\alpha}\nabla_{\alpha}\phi\,, \nn  &&
\delta(\nabla_\beta\psi^\alpha) = 2i
\bar\lambda^{\alpha\dot\alpha}\,\nabla_\beta \bar\psi_{\dot\alpha}
-2i
\bar\lambda_{\beta\dot\alpha}\,\bar\nabla^{\dot\alpha}\psi^\alpha\,,\nn
&& \delta (\bar\nabla_{\dot\alpha}\psi^{\beta}) = 2i
\bar\lambda^{\beta\dot\beta}\,\bar\nabla_{\dot\alpha}\bar\psi_{\dot\beta}
-2i\lambda^{\gamma}_{\dot\alpha}\,\nabla_{\gamma}\psi^\beta\,.
\label{TransfCD}
 \eea
The full $OSp(1|8)$ covariance of the system \p{Eqcova}, \p{Eqcovb}
is therefore obvious. In the manifestly $SL(4,R)$ covariant notation
it just reads \be \nabla_{\hat \alpha}\psi^{\hat\beta} =0\,, \quad
\nabla_{\hat\alpha}\phi = 0\,.\ee

The geometric meaning of Eqs. \p{Eqcova}, \p{Eqcovb} can be
further clarified, using Maurer-Cartan equations. Let us split the
general $osp(1|8)$ superalgebra valued Cartan form \p{Supforms}
into the parts $\Omega_\bot$ and $\Omega_{=}$, spanned,
respectively,  by the coset generators and those of the stability
subgroup $SL(4,R)$: \bea && \Omega = \Omega_\bot + \Omega_{=}\,,
\quad d\wedge \Omega + i\Omega \wedge \Omega = 0\; \Rightarrow \nn
&& {\cal T} \equiv d\wedge \Omega_\bot +i\Omega_{=} \wedge
\Omega_\bot + i\Omega_\bot \wedge \Omega_{=}  =
-i\left(\Omega_\bot\wedge \Omega_\bot\right) \vert_{\bot}\,, \label{tors} \\
&& {\cal R} \equiv d\wedge \Omega_{=} + i\Omega_{=}\wedge
\Omega_{=} = -i \left(\Omega_\bot \wedge \Omega_\bot\right)
\vert_{=}\,, \label{curv} \eea where $\vert_\bot$ and $\vert_{=}$
denote the restriction to the suitable coset and stability
subgroup generators. The supercoset \p{ll6} is not symmetric,
therefore both the torsion and curvature two-superforms ${\cal T}$
and ${\cal R}$ are non-vanishing. Since the parameters of the
coset are separated into the coordinates of the superspace
$R^{(10|4)} = (Y^{(\hat\alpha\hat\beta)}, \theta^{\hat\alpha})
\equiv \widetilde{Y}^M$ and Goldstone superfields given on
$\widetilde{Y}^M$, we actually deal with the
$R^{(10|4)}$-pullbacks of ${\cal T}$ and ${\cal R}\,$. The
supertorsion and supercurvature tensors can be defined as \be
{\cal T} = {\cal T}_{MN}\Omega^M\wedge \Omega^N\,, \quad {\cal R}
= {\cal R}_{MN}\Omega^M\wedge \Omega^N\,, \ee where \be \Omega^M =
\left(\Omega_P^{\alpha\dot\beta}, \Omega_Z^{\alpha\beta},
\bar\Omega^{\dot\alpha\dot\beta}_Z, \Omega^\alpha_Q,
\bar\Omega^{\dot\alpha}_Q\right)\,. \ee The considerations  based
on the $osp(1|8)$ anticommutation relations \p{llnew2.13a} and the
fact that the $sl(4,R)$ generators appear only in the mixed
anticommutators (between $S$ and $Q$ generators) show that Eqs.
\p{Eqcova} give rise to the vanishing of the supercurvature tensor
components in the pure Grassmann directions \be {\cal
R}_{\hat\alpha\hat\beta} = \left({\cal R}_{\alpha\beta}, {\cal
R}_{\dot\alpha\dot\beta}, {\cal R}_{\alpha\dot\beta}\right) = 0\,,
\ee while \p{Eqcovb} amount to the vanishing of certain
supertorsion components. Thus these equations are equivalent to
the particular zero-curvature (and zero-torsion) conditions.

The geometric nature of these simple dynamical conditions deserves further study.
In this connection, it is worth to note that the vanishing
of the covariant spinor world-supersurface
projections of the {\it vector} Cartan form of the target
superspace is the basic postulate of the embedding approach to
superbranes (see \cite{Dima} and refs. therein). Also, the
 vanishing of the world-supervolume spinor covariant derivative of
the spinor Goldstone superfield is the dynamical equation of $N=1,
4D$ supermembrane in the approach based on the concept of
partial breaking of global supersymmetry (PBGS) \cite{IvSK}.
One more relevant analogy is suggested by the fact that
  the superfield equations
of motion of some integrable supersymmetric $2D$ systems can be reformulated
as a dynamical inverse Higgs phenomenon (see e.g. \cite{IvKr}).

The manifestly $OSp(1|8)$ covariant
 formulation and the transformation laws \p{TransfCD} can provide
the convenient starting point for the search of the appropriate
manifestly covariant action and possible extension of the equations
 \p{Eqcova}, \p{Eqcovb} to the
 case with interaction. We also note that the system \p{Eqcova}, \p{Eqcovb}
contains only one derivative (spinor or bosonic) and so appears similar
 to the `unfolded'  form of the equations
for higher spin fields  characterized by the same  feature (see
\cite{ill5,rev2}).
Perhaps it could be put precisely in this form by adding some supplementary
equations which possibly
are satisfied as a consequence of \p{Eqcova},
\p{Eqcovb}.\footnote{E.I. thanks M. Vasiliev for suggesting this
possibility.}

\section{Tensorial chiral superspace: a proper setting for
higher spin $N=1$ supergravity?}
\setcounter{equation}{0}

An important problem for further study is the application of our
approach to higher spin extensions of $N=1$, $4D$ supergravity.
 In the standard conformal $N=1$
supergravity a purely geometric approach
 has been proposed by Ogievetsky and Sokatchev \cite{OS}.
The underlying $N=1$ supergravity gauge group in this approach is a group of
general
 diffeomorphisms of chiral $N=1$ superspace
  $C^{(4|2)} = (x_L^m, \theta^\alpha_L)$, which exposes the
   fundamental role of the principle of preserving $N=1$ chiral
    representations in $N=1$ supergravity.
    The question arises whether
an analog of this principle  can be formulated for
  higher-spin generalization of $N=1$ supergravity.
From the analysis of the full set of the (anti)commutation relations
of the superalgebra $osp(1|8)$ it follows that the minimal analog of
$C^{(4|2)}$ is the coset spanned by the following generators \be
\left(P_{\alpha\dot\alpha}, Z_{\alpha\beta}, F_{\beta\dot\beta},
Q_\alpha \right), \label{Chir} \ee i.e. it contains only one
holomorphic half of the tensorial central charges and, in addition,
the complex generator $F_{\beta\dot\beta}\,$. It is easy to check
that the rest of the $osp(1|8)$ generators form a complex
non-self-conjugated subalgebra, so the set of the coset parameters
associated with the generators \p{Chir}, i.e. \be C^{(11|2)} =
(x_L^{\alpha\dot\beta}, z_L^{\alpha\beta}, f_L^{\alpha\dot\beta},
\theta^\alpha_L) \equiv (\,Y_L\,)\,, \label{ChirSS} \ee is closed
under the left action of the supergroup $OSp(1|8)$ and provides a
natural generalization of $C^{(4|2)}$. Note that
$f_L^{\alpha\dot\beta}$ yield a holomorphic parametrization of the
coset $SL(4,R)/GL(2,C)$ and so are a sort of harmonic variables.
Thus $C^{(11|2)}$ can be also treated as an analytic subspace of the
`harmonic superspace' $R^{(10|4)}\times \frac{SL(4,R)}{GL(2,C)}$.

The precise realization of $OSp(1|8)$ in the coset manifold
\p{ChirSS} can be straightforwardly found and it will be discussed
in a future publication. Here we only give how the coordinates
\p{ChirSS} are related to the $R^{(10|4)}$ ones: \bea
\theta_L^\alpha &=& \theta^\alpha - 2i
f_L^{\alpha\dot\alpha}\bar\theta_{\dot\alpha}\,,\quad
x_L^{\alpha\dot\alpha} = x^{\alpha\dot\alpha}-
i\theta^\alpha\bar\theta^{\dot\alpha} -
4if^{\alpha}_{L\,\dot\alpha}\bar z^{\dot\alpha\dot\beta} -
(\bar\theta)^2f_L^{\alpha\dot\beta}\,, \nn z_L^{\alpha\beta}
&=&z^{\alpha\beta} + 4
f^\alpha_{L\;\;\dot\beta}f^\beta_{L\;\;\dot\gamma}\bar
z^{\dot\beta\dot\gamma} - 2
\theta^{(\alpha}f_L^{\beta)\dot\beta}\bar\theta_{\dot\beta}\,,
\label{Conn} \eea and how $f^{\alpha\dot\alpha}_L$ is transformed
under conformal supersymmetry \be \delta f_L^{\alpha\dot\alpha} =
i\bar\eta^{\dot\alpha}\theta^\alpha_L +
2\eta^{(\alpha}\theta^{\beta)}_Lf^{\dot\alpha}_{L\beta} +
(\eta\cdot\theta_L) f_L^{\alpha\dot\alpha}\,.\ee

It is interesting to inquire whether some higher-spin
 dynamics can be associated with superfields given on \p{ChirSS} as an alternative
 to eqs. \p{EQ} and what is the theory enjoying invariance under general
 diffeomorphisms of $C^{(11|2)}$ (the higher spin analog of $N=1, 4D$ conformal
 supergravity in the Ogievetsky-Sokatchev formulation?) Leaving the complete analysis of
 these issues for the future, we give here the $OSp(1|8)$ invariant tensorial superspace
 analogs of the standard kinetic and potential terms of $N=1, 4D$ chiral
 superfields. Defining the integration measures in the central and
 chiral superspaces
 \be
 \mu = d^4xd^6zd^4\theta d^4f_Ld^4f_R\,, \quad \mu_L =
 d^4x_ld^3z_Ld^2\theta_Ld^4f_L\, \quad \left(f_R^{\dot\alpha\alpha}
 \equiv \overline{(f_L^{\alpha\dot\alpha})}\right)\,,
 \label{measure}
 \ee
 one can show that they transform as
 \be
 \delta \mu = 8 \left[(\eta\cdot\theta) + (\bar\eta\cdot\bar\theta) +
 2i\left(\eta_\alpha\bar\theta_{\dot\alpha}f_L^{\alpha\dot\alpha}
 +\bar\eta_{\dot\alpha}\theta_{\alpha}f_R^{\alpha\dot\alpha}\right)\right]\mu\,,
 \quad \delta \mu_L = 12 (\eta\cdot\theta_L)\,\mu_L\,.
 \ee
 Now the $OSp(1|8)$ invariant kinetic term of the superfield $\Phi(Y_L)$ is
 uniquely defined to be
 \be
 S_{kin} \sim \int \mu\, \Phi(Y_L)\bar\Phi(Y_R)\,,
 \quad \left(Y_R = \overline{(Y_L)}\right)\label{Kin}
 \ee
 where $\Phi$ is transformed as \be \delta \Phi = - 8(\eta\cdot
 \theta_L)\,\Phi\,. \ee

The $OSp(1|8)$ invariant potential term of $\Phi(Y_L)$ is also
unique \be S_{pot} = \int \mu_L \Phi^{\frac32} +
\mbox{c.c.}\,.\label{Pot} \ee The component contents of these
actions and their relation to the higher spin theory will be
analyzed elsewhere. It still remains to give the precise meaning to
the integration over the auxiliary tensorial and $SL(4,R)/GL(2,C)$
variables.

\section{Further developments}
\setcounter{equation}{0}

In this paper we did show that nonlinear realizations of the
generalized $4D$ superconformal group $OSp(1|8)$ provide a natural
framework for treating massless higher spins in the 10-dimensional
space with tensorial coordinates. The superfield equation
encompassing all free equations for integer and half-integer spins
\cite{ill3}-\cite{ill6} was derived on a geometric ground as the
condition of vanishing of the covariant spinor projections of some
basic Cartan one-forms on the supercoset \p{ll6}. Via
Maurer-Cartan equations, this dynamical equation
implies vanishing of the supercurvature tensor along the pure odd
directions in the tensorial superspace $R^{(10|4)}\,$. The basic
scalar superfield has an intrinsic origin in the considered nonlinear
realizations framework as the coset parameter
(Goldstone superfield) associated with the spontaneously broken
dilatations. We also generalized the important notion of chirality
to the case of tensorial superspaces (with the ultimate aim to apply
this to higher spin supergravity) and constructed first examples
of the $OSp(1|8)$ invariant off-shell actions for the tensorial
chiral superfields. These actions are expected to describe a
non-trivial self-interaction of higher spins.

Besides suggesting these new insights into the theory of the
massless higher spins, the nonlinear realization approach offers
some other possibilities which we are planning to study elsewhere.
Below we list some of them.

{\bf i) AdS higher spin theories}. In order to gain massive higher
spin theories in the coset framework, e.g.
  on the AdS background,
one should pass to the curved standard and tensorial translations which belong
to the following closed set of generators.
 For example, for the bosonic $Sp(8)$ subgroup of $OSp(1|8)$ we get
\bea &&\hat P_{\alpha\dot\alpha} = P_{\alpha\dot\alpha} + m^2
K_{\alpha\dot\alpha}\,, \; \hat Z_{\alpha\beta} = Z_{\alpha\beta} +
m^2\widetilde{Z}_{\alpha\beta}\,, \;
\hat{\overline{Z}}_{\dot\alpha\dot\beta} =
\overline{Z}_{\dot\alpha\dot\beta} +m^2
\overline{\widetilde{Z}}_{\dot\alpha\dot\beta} \,,\nn &&
X_{\alpha\dot\beta} = F_{\alpha\dot\beta} + \bar
F_{\alpha\dot\beta}\,, \quad L_{\alpha\beta}, \;
\overline{L}_{\dot\alpha\dot\beta}\,, \label{pm} \eea where $m$ is a
contraction parameter having the dimension of mass (inverse AdS
radius). Introducing the coordinates just for the curved translation
generators,\footnote{The set \p{pm} is a sum of two isomorphic
algebras $sp(4) \sim o(2,3)\,$. The coordinates associated with the
curved translations
 parametrize the symmetric coset $Sp(4)\times Sp(4)/Sp(4)_{diag}\,$
  (see e.g. \cite{A2}).}
constructing the corresponding Cartan forms and imposing on them
the appropriate covariant dynamical conditions, we should obtain
the counterpart of Eqs. \p{EQ} for free higher-spin fields on
AdS$_4$ background \cite{A2,ill6}. These equations, as they stand,
are presumably related to the presented here massless ones \p{EQ}
via
 the  generalized Weyl transformation defined in \cite{A,A2}. Interaction terms
 should break full $OSp(1|8)$ symmetry and, hence, the conformal equivalence
 of the AdS$_4$ and flat cases.

{\bf ii) Tensorial analog of AdS$_{5}$ branes}. One can consider a
possible relation of the nonlinear realizations of $OSp(1|8)$ to
some AdS brane-like objects with the tensorial space as the
worldvolume. We shall limit our discussion to the bosonic group
$Sp(8)\,$. Of relevance for us will be a subgroup of $Sp(8)$
generated by the following set of generators \bea &&\hat
P_{\alpha\dot\alpha} = P_{\alpha\dot\alpha} - m^2
K_{\alpha\dot\alpha}\,, \; \hat Z_{\alpha\beta} = Z_{\alpha\beta} -
m^2\widetilde{Z}_{\alpha\beta}\,, \;
\hat{\overline{Z}}_{\dot\alpha\dot\beta} =
\overline{Z}_{\dot\alpha\dot\beta} -m^2
\overline{\widetilde{Z}}_{\dot\alpha\dot\beta} \,,\nn &&
X_{\alpha\dot\beta} = F_{\alpha\dot\beta} + \bar
F_{\alpha\dot\beta}\,, \quad L_{\alpha\beta}, \;
\overline{L}_{\dot\alpha\dot\beta}\,. \label{pm1} \eea
The group generated by (\ref{pm1})
  contains $SO(1,4)
   \propto (P_{\alpha\dot\beta}
    - m^2 K_{\alpha\dot\beta},
    L_{\alpha\beta}, \bar L_{\dot\alpha\dot\beta})$
as a subgroup and describes an
 extension of  $SO(1,4)$ by the tensorial nonlinear translations generated by
 $\hat{{Z}}_{\alpha\beta}$, $\overline{\hat{Z}}_{\dot\alpha\dot\beta}$
  and $X_{\alpha\dot{\beta}}\,$.

  In $4D$ case we can parametrize the AdS$_5$ coset
  $\frac{SO(2,4)}{SO(1,4)}$ by the coordinates $x^{\alpha\dot{\beta}}$
   and dilaton  $\phi$ \cite{Solv}, and obtain the description of AdS$_5$ $3$-brane \cite{BIK}.
In the $Sp(8)$ case
the true analog of AdS$_5$ is just the coset of $Sp(8)$ over
the subgroup generated by \p{pm1}.
This $Sp(8)$ coset manifold contains AdS$_5$ as a subspace,
 but it is much larger,
because the full set of the coset generators  is the following
\be P_{\alpha\dot\alpha},
Z_{\alpha\beta}, \bar Z_{\dot\alpha\dot\beta}, D, A,
G_{\gamma\dot\gamma} = i(F_{\gamma\dot\gamma} - \bar F_{\gamma\dot\gamma})\,.\label{Gen}
\ee
It contains
\\
- 10-dimensional extended space-time manifold $R^{10} =
(x^{\alpha\dot\alpha}, z^{\alpha \beta}, \bar{z}^{\dot{\alpha}\dot{\beta}})$
 associated with the generators $(P_{\alpha\dot\alpha},
 Z_{\alpha \beta}, \bar{Z}_{\dot{\alpha}\dot{\beta}})\,$.
 It is supplemented by dilaton.
\\
   - additional 5 dimensions generated by $A$ and
   $G_{\alpha \dot{\beta}}\,$.

We see that the $Sp(8)$ analog of the AdS$_5$ 3-brane with
$R^{10}$ as the worldvolume should involve besides Goldstone
dilation field also further five transverse coordinates: one
pseudoscalar coordinate  generated by $A$
  and a real vector one
associated with $G_{\alpha\dot\alpha}$.
 It is  tempting to  describe such an
  exotic brane-like object (and its superextension related to $OSp(1|8)$)
 and to see how it is related to the higher-spin theories.

{\bf iii) Towards higher dimensions}. In this work we considered the
$4D$
 case for simplicity. The generalization
of our approach to $D>4$ implies the application of appropriate
nonlinear coset realizations of the generalized $11\geq D
> 4$ superconformal
 algebras described
by suitable real forms of $OSp(1|2^k)$ ($6 \geq k > 3$).
  Because the higher spin
theories in diverse dimensions are intensively studied
 (see \cite{rev2} and refs. therein, as well as \cite{ill6} - in the context
 of tensorial superspaces), such generalization should be also investigated.

\subsubsection*{Acknowledgments}
The authors would like to thank Dmitri Sorokin for numerous valuable
discussions and comments and Misha Vasiliev for interest in the work
and discussion. E.I. acknowledges a partial support from RFBR
grants, projects  No 03-02-17440 and No 04-02-04002, NATO grant
PST.GLG.980302, the grant INTAS-00-00254, the DFG grant No.436 RUS
113/669-02, and a grant of the Heisenberg-Landau program.

\end{document}